RESEARCH ARTICLE

# A Computational Study on the Relation between Resting Heart Rate and Atrial Fibrillation Hemodynamics under Exercise

Matteo Anselmino[1]*, Stefania Scarsoglio[2], Andrea Saglietto[1], Fiorenzo Gaita[1], Luca Ridolfi[3]

1 Division of Cardiology, Department of Medical Sciences, "Città della Salute e della Scienza" Hospital, University of Turin, Turin, Italy, 2 DIMEAS -Department of Mechanical and Aerospace Engineering-, Politecnico di Torino, Turin, Italy, 3 DIATI -Department of Environmental, Land and Infrastructure Engineering-, Politecnico di Torino, Turin, Italy

* matteo.anselmino@unito.it

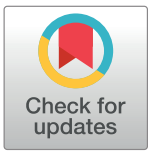

**Data Availability Statement:** All data have been made publicly available from the Figshare repository at the following URL: https://figshare.com/s/4e43bb2df0f60278ef31 (DOI: https://dx.doi.org/10.6084/m9.figshare.4203255).

**Funding:** The authors received no specific funding for this work.

**Competing Interests:** The authors have declared that no competing interests exist.

## Abstract

### Aims

Clinical data indicating a heart rate (HR) target during rate control therapy for permanent atrial fibrillation (AF) and assessing its eventual relationship with reduced exercise tolerance are lacking. The present study aims at investigating the impact of resting HR on the hemodynamic response to exercise in permanent AF patients by means of a computational cardiovascular model.

### Methods

The AF lumped-parameter model was run to simulate resting (1 Metabolic Equivalent of Task—MET) and various exercise conditions (4 METs: brisk walking; 6 METs: skiing; 8 METs: running), considering different resting HR (70 bpm for the slower resting HR—SHR—simulations, and 100 bpm for the higher resting HR—HHR—simulations). To compare relative variations of cardiovascular variables upon exertion, the variation comparative index (VCI)—the absolute variation between the exercise and the resting values in SHR simulations referred to the absolute variation in HHR simulations—was calculated at each exercise grade ($VCI_4$, $VCI_6$ and $VCI_8$).

### Results

Pulmonary venous pressure underwent a greater increase in HHR compared to SHR simulations ($VCI_4 = 0.71$, $VCI_6 = 0.73$ and $VCI_8 = 0.77$), while for systemic arterial pressure the opposite is true ($VCI_4 = 1.15$, $VCI_6 = 1.36$, $VCI_8 = 1.56$).

### Conclusions

The computational findings suggest that a slower, with respect to a higher resting HR, might be preferable in permanent AF patients, since during exercise pulmonary venous pressure undergoes a slighter increase and systemic blood pressure reveals a more appropriate increase.

## Introduction

Atrial fibrillation (AF), the most common cardiac arrhythmia, is a major healthcare burden, whose prevalence is constantly rising. In case of permanent AF, a rate control approach combined with an oral anticoagulant therapy (OAT) is generally adopted[1] to decrease the risk of thromboembolic events, reduce symptoms and improve quality of life (QoL). However, clinical evidence indicating a clear-cut target heart rate (HR) is lacking. In fact, one of the few studies concerning this topic, the RACE II[2] trial, suggested that a lenient rate control approach (target resting HR < 110 bpm) might not be inferior to a strict rate control approach (target resting HR < 80 bpm) in terms of cardiovascular outcomes. However, this trial has been widely debated[3], requiring further clinical evidence. In the meantime, recent cardiovascular computational models[4] suggest, instead, that lower HRs during permanent AF relate to improved hemodynamic parameters, cardiac efficiency, and lower oxygen consumption[5].

Reduced exercise tolerance is, together with chest pain, palpitations and shortness of breath, one of the major symptoms of AF. The decline in exercise capacity during AF is typically on the order of 15% to 20%[6, 7]. Drop in exercise performance during lone AF was found regardless of adequate ventricular rate control [8, 9]. Short- and long-term improvements in exercise response were observed in patients with AF after cardioversion[10]. Recently, a weak positive correlation between resting heart rate and exercise capacity was found during AF [11]. On the contrary, baseline heart rate, together with age and ratio of early mitral inflow velocity to mitral annular velocity, was recognized as an independent predictor of exercise intolerance in AF patients[12].

This topic is still widely debated, and the relationship between AF and exercise tolerance remains uncertain. The underlying mechanisms are not clear, but most probably rapid and irregular rate plays a key role, together with the consequent reduced passive ventricular filling in early diastole. Being likely that a greater increase in HR (given the same amount of exertion) is related to a worse tolerance[13], it is still questioned whether and how the resting HR in permanent AF patients influences the cardiovascular system response to physical exercise. For this reason, also considering the growing impact of computational studies into cardiac translational research[14, 15], a computational approach to explore the relationship between AF and exercise capacity is here proposed.

Starting from a lumped mathematical algorithm [4, 5, 16, 17] validated in resting conditions, we designed a model to simulate cardiovascular system response during AF and physical exertion, given different resting HRs: 70 bpm for the slower resting HR (SHR simulations) and 100 bpm for the higher resting HR (HHR simulations). Using Metabolic Equivalent of Task (MET)[18] several exercise conditions were modelled and simulated (1 MET: resting condition, 4 METs: brisk walking; 6 METs: skiing; 8 METs: running). To compare relative variations of cardiovascular variables upon exertion, we evaluated at each exercise grade the variation comparative index (VCI). VCI is the absolute variation between the exercise and the resting values in SHR simulations referred to the absolute variation in HHR simulations. We therefore compared relative changes in several hemodynamic variables elicited by the same amount of exertion, quantifying how resting HR in permanent AF might influence cardiovascular performance. In our opinion, the present computational model brings new insights into the relationship between resting heart rate and hemodynamic response during AF under different grades of exertion.

## Materials and Methods

### Lumped-model

The hereby adopted mathematical model is able to measure the effects of AF on the cardiovascular system and it has extensively been validated in resting conditions through systematic comparison with more than thirty clinical studies providing AF hemodynamic data[4]. The present modeling, first proposed by Korakianitis and Shi[19], is based on a lumped parameterized approach of the beating heart together with the systemic and pulmonary circuits, so that both cardiac and circulatory systems are described in terms of electrical components, such as resistances, inductances and compliances. The resulting ordinary differential system is expressed by means of pressures, P, volumes, V, and flow rates, Q, for each heart chamber and vascular section, and is solved through an adaptive multistep numerical scheme implemented in Matlab (for details, please refer to Scarsoglio et al., 2014).

### Variables definition

Left heart dynamics were evaluated in terms of volume [ml] and pressure [mmHg], also considering end-diastolic (ed) and end-systolic (es) values: left atrial pressure ($P_{laed}$), left atrial volume ($V_{laed}$), left ventricular pressure ($P_{lved}$), left ventricular volume ($V_{lved}$). End-systolic values refer to the instant defined by the closure of the aortic valve, while end-diastolic values correspond to the closure of the mitral valve. Left ventricular performance was evaluated through the following hemodynamic parameters: stroke volume, SV = $V_{lved}$—$V_{lves}$ [ml], ejection fraction, EF = SV/$V_{lved}$ x 100% [%], stroke work, SW/min [J/min], measured as the area within the left ventricle pressure-volume loop per beat, and cardiac output, CO = SV x HR [l/min]. Oxygen consumption was also estimated through the tension time index per minute, TTI/min = $P_{lv}$ x RR x HR [mmHg s/min]. In addition, systemic arterial ($P_{sas}$, $P_{sas,syst}$, $P_{sas,dias}$), pulmonary arterial ($P_{pas}$, $P_{pas,syst}$, $P_{pas,dias}$) and venous ($P_{pvn}$) pressures [mmHg] were evaluated.

### Exercise simulation parameters

In terms of "metabolic equivalent of task" (MET), resting simulation corresponds to 1 MET, while exercise simulations were designed to resemble, respectively, a light effort (4 METs—e.g. brisk walking), a moderate effort (6 METs—e.g. skiing) and a vigorous effort (8 METs—e.g. running). Starting from 1 MET (see S1 Table), exercise parameters were tuned (Table 1) according to the physiological response of the cardiovascular system. Based on hemodynamic data available in literature [20–24], parameters were modified and simulations were run in SR to test performance compared to literature data during different levels of exercise [20, 25]:

- maximum left ventricular ($E_{lv,max}$) and maximum right ventricular elastance ($E_{rv,max}$) were increased, given the enhanced cardiac inotropism[20, 21];

- systemic arterial compliances ($C_{sas}$ and $C_{sat}$) were diminished, due to the increase in vasoconstrictor tone, in blood pressure and in HR associated with exercise[22, 23];

- systemic arteriolar and capillary resistances ($R_{sar}$ and $R_{scp}$) were reduced, the former due to massive vasodilation in skeletal muscles, the latter due to capillary recruitment in the same district[20];

**Table 1. Cardiovascular parameters modified for the tuning of the model to simulate physiological response of the cardiovascular system to exertion (4, 6 and 8 METs).**

| Parameters | 1 MET | 4 METs | 6 METs | 8 METs |
|---|---|---|---|---|
| $E_{lv,max}$ [mmHg/ml] | 2.5 | 6 | 8 | 10 |
| $E_{rv,max}$ [mmHg/ml] | 1.15 | 2.3 | 3 | 3.5 |
| $C_{sas}$ [ml/mmHg] | 0.064 | 0.0576 | 0.0512 | 0.0448 |
| $C_{sat}$ [ml/mmHg] | 1.28 | 1.152 | 1.024 | 0.896 |
| $R_{sar}$ [mmHg s/ml] | 0.44 | 0.2948 | 0.2332 | 0.1936 |
| $R_{scp}$ [mmHg s/ml] | 0.4576 | 0.3066 | 0.2425 | 0.2013 |
| $C_{svn}$ [ml/mmHg] | 20.5 | 14 | 9.5 | 5 |
| $R_{pcp}$ [mmHg s/ml] | 0.0385 | 0.0258 | 0.0204 | 0.0181 |

$C_{sas}$, systemic aortic sinus compliance; $C_{sat}$, systemic arterial compliance; $C_{svn}$, systemic venous resistance; $E_{lv,max}$, maximum left ventricular elastance; $E_{rv,max}$, maximum right ventricular elastance; $R_{pcp}$, pulmonary capillary resistance; $R_{sar}$, systemic arteriolar resistance; $R_{scp}$, systemic capillary resistance.

doi:10.1371/journal.pone.0169967.t001

- systemic venous compliance ($C_{svn}$) was reduced, accounting for the enhanced venous vascular tone[20];
- pulmonary capillary resistance ($R_{pcp}$) was reduced, due to capillary recruitment especially in the superior portion of the lung[24].

To validate the tuning of these parameters, four simulations (1 MET, 4 METs, 6 METs and 8 METs) were first run in sinus rhythm (SR). To take into account the enhanced cardiac chronotropism present during exercise, simulations were set at progressively faster HRs (1 MET: 70 bpm; 4 METs: 90 bpm; 6 METs: 110 bpm; 8 METs: 130 bpm). The results of these SR testing simulations (Table 2) proved, in fact, to agree with *in vivo* measurements during different levels of exercise in sinus rhythm subjects[20, 25]. In particular, for increasing level of exercise (from resting to vigorous exercise), the tuning results showed overall similar trends to *in vivo* data (systolic arterial pressure: +43% vs. +19–42%; diastolic arterial pressure: +19% vs. +10–29%; cardiac output: +160% vs. +77–167%).

## AF simulations

In order to mimic the absence of atrial systole, both atria are imposed as passive (in S1 Table, $E_{la} = E_{la,min}$ and $E_{ra} = E_{ra,min}$ through the whole cardiac period). Atrial compliance is also supposed to reduce during AF, both acutely due to asynchronous atrial contractions[26] and chronically due to fibrosis[27]. However, since no further definitive data concerning atrial compliance variation in AF are available, we presently omitted this aspect. Starting from two different resting HR, slower heart rate (SHR, HR = 70 bpm at rest) and higher heart rate (HHR, HR = 100 bpm at rest), exercise simulations were designed accounting for a progressive HR increase. SHR exercise simulations were run at 90 bpm (4 METs), 110 bpm (6 METs) and 130 bpm (8 METs), while HHR exercise simulations at 120 bpm (4 METs), 140 bpm (6 METs) and 160 bpm (8 METs), respectively. Given the lack of clear data concerning ventricular dysfunction during AF under exercise[4], the reduced LV inotropy was not modelled and a normal left ventricular contractility was assumed as baseline in both resting conditions. The irregularity of AF beating is simulated by extracting RR intervals from an exponentially modified Gaussian (EMG) distribution, which is the most common RR distribution recorded during AF[4]. The adopted RR beats were obtained by the superposition of two statistically independent times, extracted from a correlated pink Gaussian distribution and an

**Table 2. Steady state values of the computed cardiovascular variables in SR testing simulations with constant RR.**

| Variables | 1 MET (70 bpm) | 4 METs (90 bpm) | 6 METs (110 bpm) | 8 MET (130 bpm) |
|-----------|----------------|-----------------|------------------|-----------------|
| $P_{laed}$ [mmHg] | 9.13 | 9.17 | 10.29 | 12.48 |
| $P_{lved}$ [mmHg] | 10.71 | 15.55 | 23.35 | 35.35 |
| $V_{laed}$ [ml] | 58.19 | 58.50 | 65.95 | 80.55 |
| $V_{lved}$ [ml] | 130.91 | 125.29 | 127.83 | 131.71 |
| $P_{sas}$ [mmHg] | 97.13 | 111.63 | 120.71 | 131.19 |
| $P_{sas,dias}$ [mmHg] | 74.44 | 80.96 | 84.66 | 88.58 |
| $P_{sas,syst}$ [mmHg] | 120.39 | 141.31 | 155.52 | 172.60 |
| $P_{pas}$ [mmHg] | 17.11 | 19.82 | 22.90 | 27.49 |
| $P_{pas,dias}$ [mmHg] | 11.12 | 11.85 | 13.88 | 17.41 |
| $P_{pas,syst}$ [mmHg] | 25.92 | 31.38 | 35.45 | 40.74 |
| $P_{pvn}$ [mmHg] | 10.35 | 10.73 | 12.08 | 14.53 |
| SV [ml] | 74.25 | 92.83 | 98.55 | 103.85 |
| EF [%] | 56.72 | 74.09 | 77.08 | 78.85 |
| CO [l/min] | 5.20 | 8.35 | 10.84 | 13.50 |
| TTI/min [mmHg s/min] | 2,615 | 3,580 | 4,358 | 5,230 |
| SW/min [J/min] | 69.17 | 139.60 | 206.13 | 293.11 |

CO, cardiac output; EF, ejection fraction; $P_{laed}$, left atrium end-diastolic pressure; $P_{lved}$, left ventricular end-diastolic pressure; $P_{pas}$, mean pulmonary arterial pressure; $P_{pas,dias}$, diastolic pulmonary arterial pressure; $P_{pas,syst}$, systolic pulmonary arterial pressure; $P_{pvn}$, pulmonary vein pressure; $P_{sas}$, mean systemic arterial pressure; $P_{sas,dias}$, diastolic systemic arterial pressure; $P_{sas,syst}$, systolic systemic arterial pressure; SV, stroke volume; SW/min, stroke work per minute; TTI/min, tension time index; $V_{laed}$, left atrium end-diastolic volume; $V_{lved}$, left ventricular end-diastolic volume.

uncorrelated exponential distribution. By varying the HR from 70 to 160 bpm, every distribution is built keeping the coefficient of variation, cv = $\sigma/\mu$ ($\mu$ = 60/HR: mean of the RR distribution, $\sigma$: standard deviation), equal to 0.24, which is the typical value observed during AF[28]. The rate parameter, $\gamma$, related to the exponential term of the resulting EMG distribution, is taken as function of the mean HR and was estimated by means of the RR recordings of the MIT-BIH Long-Term AF Database[29, 30]. Details of the $\gamma$-HR fitting are given in Fig 1. The resulting RR probability distribution functions are displayed in Fig 2, while mean and standard deviation values are offered in Table 3.

For each HR, the model equations are integrated in time until the first two statistics (mean and standard deviation) of the cardiovascular parameters remain insensitive to further extensions of the temporal computational domain. 5,000 cardiac periods guarantee the statistical stationarity of the modeling results to be reached. Thus, all the variables presented in the following sections are intended as statistical results averaged over 5,000 cycles.

## Performance comparison

To compare relative variations of the computational results, "variation comparative index" (VCI) was used. This index, evaluated for every cardiovascular variable at each exercise intensity ($VCI_4$, $VCI_6$ and $VCI_8$), is defined as the absolute variation between the exercise and the resting values in SHR simulations divided by the absolute variation in HHR simulations. Thus, for a general hemodynamic variable, q, VCI at a specific exercise intensity (ex = 4, 6, 8 METs), can be expressed as follows: $VCI_{ex} = (q_{ex}-q_{res})_{SHR} / (q_{ex}-q_{res})_{HHR}$, where the subscript *res* indicates the resting value (1 MET) of q.

Through VCI, we compare how the resting HR during permanent AF influences the exercise-induced modifications of cardiovascular variables. A VCI > 1 indicates that the studied variable undergoes a greater variation in SHR exercise simulation, while a VCI < 1 means that the greatest variation occurs in HHR exercise simulation. A variation was considered relevant when VCI ≥ 1.15 or ≤ 0.85, given the assumption that ±15% represents the estimated maximum variation between instrumental and operator-dependent measurements.

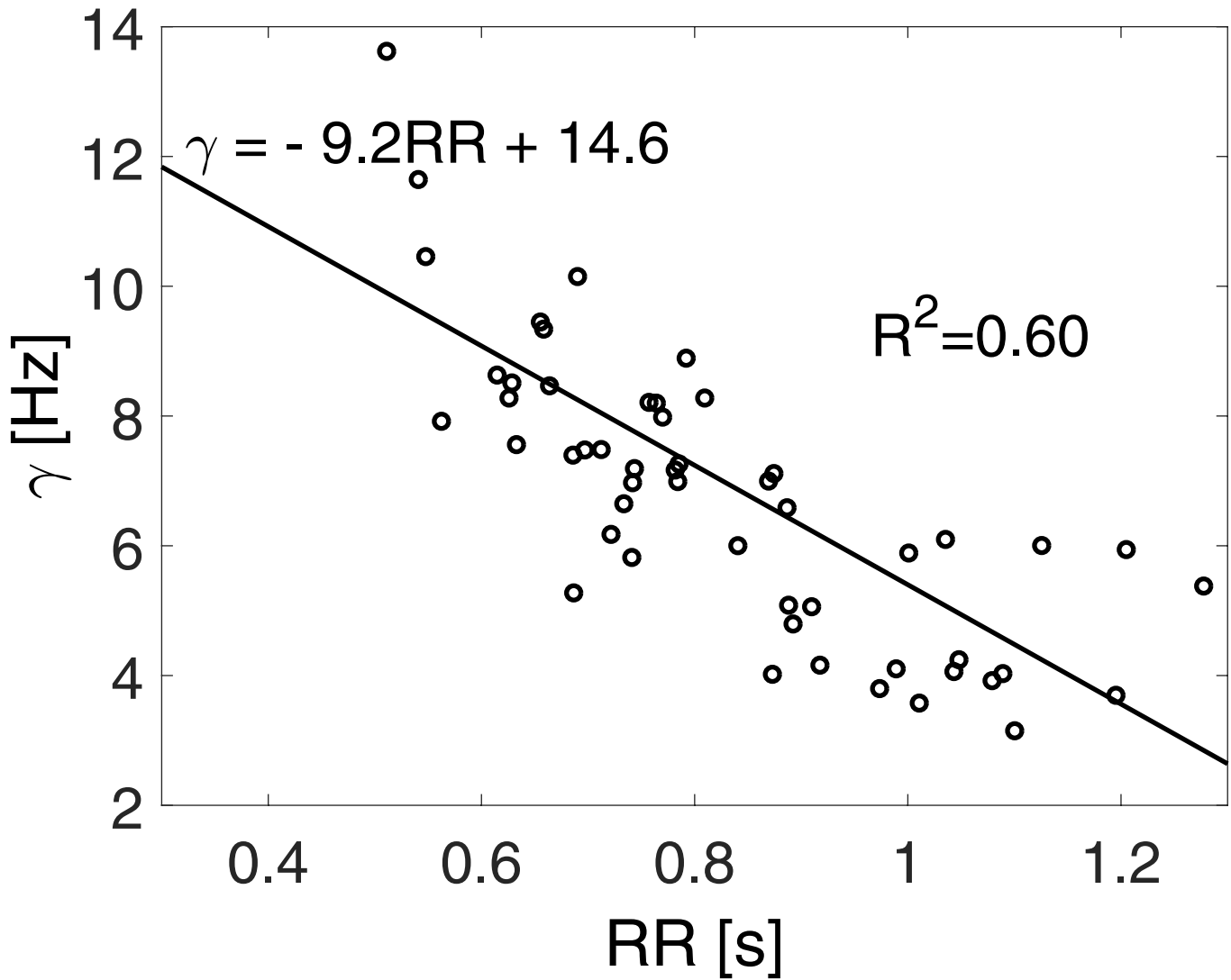

**Fig 1. Rate parameter, γ, as a function of the mean RR interval.** The Long-Term AF Database [29, 30] contains 84 long-term ECG recordings of subjects with paroxysmal or sustained AF. Record durations vary but cover typically 24 to 25 hours. Among the 84 recordings we focus on those with unimodal distributions having at maximum 10 minutes of other arrhythmias (e.g. sinus bradycardia, idioventricular rhythm, ventricular bigeminy, etc), thereby selecting 26 recordings. For each of these series, two windows of 10,000 beatings have been extracted, allowing the evaluation of 52 power spectrum density functions. It has been observed[31] that $\gamma = S_\infty^{-1/2}$, where $S_\infty$ is the flat power spectrum density at the highest frequencies. The rate parameter, γ, has been here evaluated over the 100 highest frequencies for each of the 52 RR series selected. γ values are reported through open circles as function of the mean RR of each chosen window. A linear least squares fitting of the data is introduced (solid line) together with the coefficient of determination, $R^2 = 0.60$. The resulting relation γ = -9.2RR + 14.6 is adopted to obtain γ values for the different HR targets (specific values are reported in Table 3).

## Results

The lumped model was run to compute several cardiovascular variables for SHR and HHR simulations. Mean values and standard deviation values of the computed variables over 5,000 cardiac cycles are reported in Table 4, while VCIs calculated for every cardiovascular variable at each exercise intensity ($VCI_4$, $VCI_6$ and $VCI_8$) are reported in Table 5 and displayed in Fig 3.

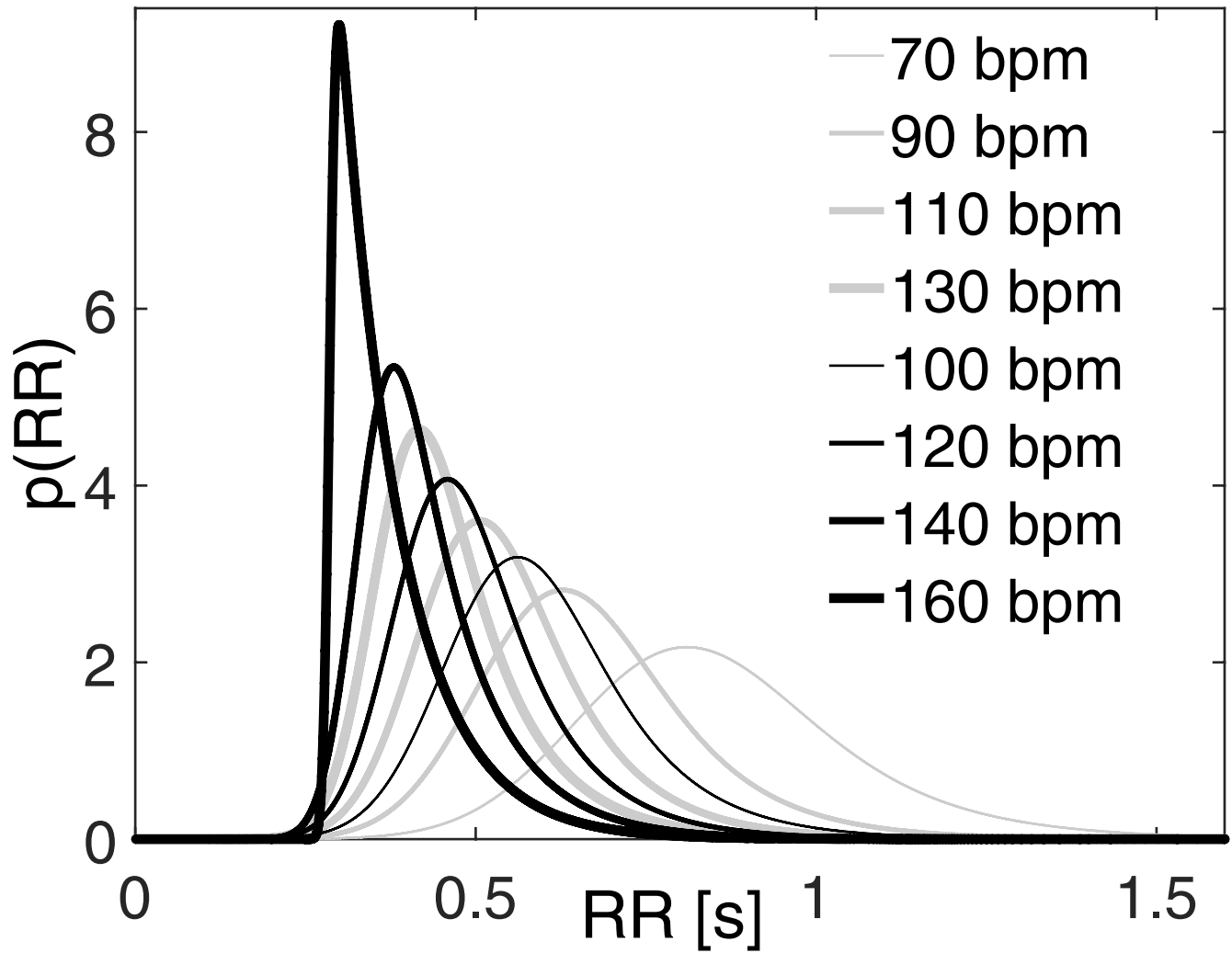

**Fig 2. Probability distribution functions of the RR beating for the different AF simulations.** Mean, µ, standard deviation, σ, and rate parameter, γ, values of each configuration are reported in Table 3.

Simulation results are here mainly presented by means of VCIs (Table 5) rather than specific hemodynamic values, as the former is more suitable to synthetize whether significant changes occur during SHR (VCI > 1) or HHR (VCI < 1) regimes. The systemic arterial pressure variations ($P_{sas}$, $P_{sas,dias}$, $P_{sas,syst}$), as well as SV, are significantly greater during SHR

**Table 3. Parameters of the RR series extracted for the AF simulations: µ (mean value), σ (standard deviation), γ (rate parameter).**

| | SHR AF simulations | | | | HHR AF simulations | | | |
|---|---|---|---|---|---|---|---|---|
| | 70 bpm | 90 bpm | 110 bpm | 130 bpm | 100 bpm | 120 bpm | 140 bpm | 160 bpm |
| µ [s] | 0.8571 | 0.6667 | 0.5455 | 0.4615 | 0.6 | 0.5 | 0.4286 | 0.3750 |
| σ [s] | 0.2057 | 0.16 | 0.1309 | 0.1108 | 0.1440 | 0.12 | 0.1029 | 0.09 |
| γ [Hz] | 6.7143 | 8.4667 | 9.5818 | 10.3538 | 9.08 | 10 | 10.6571 | 11.15 |

Table 4. Mean values and standard deviations of the computed cardiovascular variables in AF simulations.

| Variables | SHR AF simulations | | | | HHR AF simulations | | | |
|---|---|---|---|---|---|---|---|---|
| | 1 MET (70 bpm) | 4 METs (90 bpm) | 6 METs (110bpm) | 8 METs (130 bpm) | 1 MET (100 bpm) | 4 METs (120 bpm) | 6 METs (140 bpm) | 8 METs (160 bpm) |
| $P_{laed}$ [mmHg] | 10.34 ± 0.15 | 10.43 ± 0.22 | 11.39 ± 0.26 | 13.18 ± 0.52 | 10.14 ± 0.11 | 10.32 ± 0.29 | 11.68 ± 0.48 | 14.09 ± 0.67 |
| $P_{lved}$ [mmHg] | 17.25 ± 0.94 | 23.21 ± 1.77 | 29.23 ± 2.37 | 37.82 ± 3.39 | 18.72 ± 1.16 | 25.45 ± 1.98 | 32.60 ± 2.62 | 42.87 ± 3.10 |
| $V_{laed}$ [ml] | 66.27 ± 1.01 | 66.88 ± 1.44 | 73.28 ± 1.72 | 85.19 ± 3.44 | 64.92 ± 0.76 | 66.15 ± 1.95 | 75.23 ± 3.19 | 91.27 ± 4.49 |
| $V_{lved}$ [ml] | 125.17 ± 4.94 | 118.62 ± 9.27 | 121.58 ± 13.19 | 126.86 ± 20.42 | 114.63 ± 6.76 | 104.90 ± 12.02 | 105.70 ± 17.63 | 106.39 ± 24.62 |
| $P_{sas}$ [mmHg] | 92.26 ± 15.40 | 105.69 ± 20.46 | 114.67 ± 24.02 | 126.17 ± 29.33 | 102.52 ± 12.28 | 114.16 ± 17.02 | 119.04 ± 20.75 | 124.24 ± 25.76 |
| $P_{sas,dias}$ [mmHg] | 72.12 ± 8.75 | 78.54 ± 10.38 | 82.21 ± 10.62 | 86.90 ± 11.66 | 85.76 ± 6.65 | 90.59 ± 8.13 | 90.98 ± 9.40 | 91.46 ± 11.73 |
| $P_{sas,syst}$ [mmHg] | 115.83 ± 4.78 | 135.39 ± 4.70 | 149.39 ± 4.58 | 167.69 ± 9.15 | 121.77 ± 2.75 | 138.15 ± 4.03 | 147.02 ± 7.29 | 156.91 ± 13.30 |
| $P_{pas}$ [mmHg] | 16.89 ± 4.84 | 19.50 ± 6.24 | 22.45 ± 6.98 | 26.87 ± 7.75 | 17.44 ± 4.05 | 20.19 ± 5.43 | 23.29 ± 6.09 | 27.62 ± 6.55 |
| $P_{pas,dias}$ [mmHg] | 11.35 ± 0.81 | 12.23 ± 1.17 | 14.17 ± 1.42 | 17.46 ± 1.74 | 12.32 ± 0.96 | 13.42 ± 1.27 | 15.68 ± 1.42 | 19.24 ± 1.52 |
| $P_{pas,syst}$ [mmHg] | 25.48 ± 1.36 | 30.63 ± 1.89 | 34.64 ± 2.03 | 39.95 ± 2.35 | 24.08 ± 1.06 | 29.14 ± 1.55 | 33.15 ± 1.91 | 37.92 ± 2.40 |
| $P_{pvn}$ [mmHg] | 10.53 ± 0.71 | 10.97 ± 0.96 | 12.26 ± 1.06 | 14.51 ± 1.22 | 10.28 ± 0.57 | 10.90 ± 0.82 | 12.65 ± 0.97 | 15.45 ± 1.12 |
| SV [ml] | 70.15 ± 7.26 | 86.92 ± 10.29 | 92.99 ± 13.46 | 99.60 ± 19.90 | 54.98 ± 7.89 | 70.88 ± 12.12 | 75.95 ± 17.03 | 79.11 ± 23.39 |
| EF [%] | 55.91 ± 3.89 | 73.01 ± 3.77 | 76.12 ± 3.74 | 77.88 ± 4.41 | 47.71 ± 4.51 | 67.02 ± 5.16 | 71.06 ± 5.18 | 73.28 ± 5.00 |
| CO [l/min] | 5.11 ± 1.05 | 8.19 ± 1.80 | 10.63 ± 2.42 | 13.38 ± 3.49 | 5.72 ± 1.17 | 8.86 ± 2.10 | 11.07 ± 3.13 | 13.19 ± 4.54 |
| TTI/min [mmHg s/min] | 2,572 ± 302 | 3,516 ± 446 | 4,270 ± 516 | 5,166 ± 641 | 3,064 ± 325 | 4,028 ± 448 | 4,674 ± 554 | 5,340 ± 751 |
| SW/min [J/min] | 65.94 ± 15.38 | 132.55 ± 33.28 | 196.38 ± 50.27 | 286.85 ± 85.70 | 81.97 ± 17.75 | 153.28 ± 38.52 | 210.89 ± 65.33 | 278.99 ± 113.39 |

CO, cardiac output; EF, ejection fraction; $P_{laed}$, left atrium end-diastolic pressure; $P_{lved}$, left ventricular end-diastolic pressure; $P_{pas}$, mean pulmonary arterial pressure; $P_{pas,dias}$, diastolic pulmonary arterial pressure; $P_{pas,syst}$, systolic pulmonary arterial pressure; $P_{pvn}$, mean pulmonary venous pressure; $P_{sas}$, mean systemic arterial pressure; $P_{sas,dias}$, diastolic systemic arterial pressure; $P_{sas,syst}$, systolic systemic arterial pressure; SV, stroke volume; SW/min, stroke work per minute; TTI/min, tension time index per minute; $V_{laed}$, left atrium end-diastolic volume; $V_{lved}$, left ventricular end-diastolic volume.

simulations (VCIs > 1.15). On the contrary, left atrial and ventricular variables ($P_{laed}$, $V_{laed}$, $P_{lved}$, $V_{lved}$) are more prone to substantial changes during HHR simulations (VCIs < 0.85), as well as $P_{pvn}$ and $P_{pas,dias}$. Regarding the other cardiovascular variables, such as the mechano-energetic indexes of left heart (EF, CO, TTI/min, SW/min), they do not undergo significant changes (0.85 < VCIs < 1.15).

Concerning the oxygen consumption (TTI/min) and the cardiac power (SW/min), VCIs greater than 1 do not mean that, at a fixed exercise grade, the SHR is worse than the HHR in absolute terms. In fact, if we compare mean values of TTI/min and SW/min at the same exercise grade, we observe greater values of both variables during HHR with respect to SHR (the only exception is SW/min at 8 METs). This trend is in agreement with the behavior reported in [5] for different resting HRs, where both TTI/min and SW/min increase with HR. From resting conditions to moderate effort, the increased oxygen consumption when passing from SHR to HHR is partially compensated by a corresponding increased cardiac power. However, during vigorous effort (8 METs) this compensation saturates, leading to an increased oxygen consumption which is accompanied by a lower cardiac power at HHR.

**Table 5. VCIs of the computed variables†.**

| Variables | VCI$_4$ | VCI$_6$ | VCI$_8$ |
|---|---|---|---|
| P$_{laed}$ [mmHg] | **0.49** | **0.68** | **0.72** |
| P$_{lved}$ [mmHg] | **0.89** | **0.86** | **0.85** |
| V$_{laed}$ [ml] | **0.49** | **0.68** | **0.72** |
| V$_{lved}$ [ml] | **0.67** | **0.40** | **0.21** |
| P$_{sas}$ [mmHg] | **1.15** | **1.36** | **1.56** |
| P$_{sas,dias}$ [mmHg] | **1.33** | **1.93** | **2.59** |
| P$_{sas,syst}$ [mmHg] | **1.19** | **1.33** | **1.48** |
| P$_{pas}$ [mmHg] | 0.95 | 0.95 | 0.98 |
| P$_{pas,dias}$ [mmHg] | **0.80** | **0.84** | **0.88** |
| P$_{pas,syst}$ [mmHg] | 1.02 | 1.01 | 1.05 |
| P$_{pvn}$ [mmHg] | **0.71** | **0.73** | **0.77** |
| SV [ml] | **1.05** | **1.09** | **1.22** |
| EF [%] | 0.89 | 0.87 | 0.86 |
| CO [l/min] | 0.98 | 1.03 | 1.10 |
| TTI/min [mmHg s/min] | 0.98 | 1.06 | 1.14 |
| SW/min [J/min] | 0.93 | 1.01 | 1.12 |

CO, cardiac output; EF, ejection fraction; P$_{laed}$, left atrium end-diastolic pressure; P$_{lved}$, left ventricular end-diastolic pressure; P$_{pas}$, mean pulmonary arterial pressure; P$_{pas,dias}$, diastolic pulmonary arterial pressure; P$_{pas,syst}$, systolic pulmonary arterial pressure; P$_{pvn}$, pulmonary venous pressure; P$_{sas}$, mean systemic arterial pressure; P$_{sas,dias}$, diastolic systemic arterial pressure; P$_{sas,syst}$, systolic systemic arterial pressure; SV, stroke volume; SW/min, stroke work per minute; TTI/min, tension time index per minute; V$_{laed}$, left atrium end-diastolic volume; V$_{lved}$, left ventricular end-diastolic volume.

† Variables showing at least one of the three VCIs (VCI$_4$, VCI$_6$, VCI$_8$) $\geq$ 1.15 or $\leq$ 0.85 are reported in bold.

doi:10.1371/journal.pone.0169967.t005

To test the sensitivity of the results, we have varied each of the input parameters reported in Table 1 by +5% and -5% in the two extreme conditions (1 MET and 8 METs) during SHR. Results of a selection of significant hemodynamic outputs (P$_{sas}$, P$_{pvn}$, P$_{pas,dias}$, SV) are reported in S2 Table. The general behavior remained unaltered with respect to the results obtained using parameters as in Table 1, since output variations are within 2%. Thus, the chosen input parameters of Table 1 were assumed as robust reference baseline values.

## Discussion

Among all the hemodynamic parameters analyzed in the present computational model, pulmonary venous pressure and systemic arterial pressure significantly changed upon exertion during permanent AF. Pulmonary venous pressure underwent a greater increase, while systemic arterial pressure experienced a less sustained increase in HHR compared to SHR simulations.

Despite it is generally accepted that exercise tolerance is reduced in AF patients[6], this topic is yet poorly explored. In particular, clinical data regarding a possible correlation between resting HR achieved during AF rate control therapy and exercise capacity are insufficient and controversial. Recently, a study by Kato et al.[11], designed to compare the relationship between resting HR and peak oxygen consumption (VO$_2$) between sinus rhythm and AF, demonstrated either a weak positive or no correlation between resting HR and peak VO$_2$ in AF patients. On the contrary, in a previous study by Lee et al.[12], resting HR was negatively correlated to exercise capacity in AF patients.

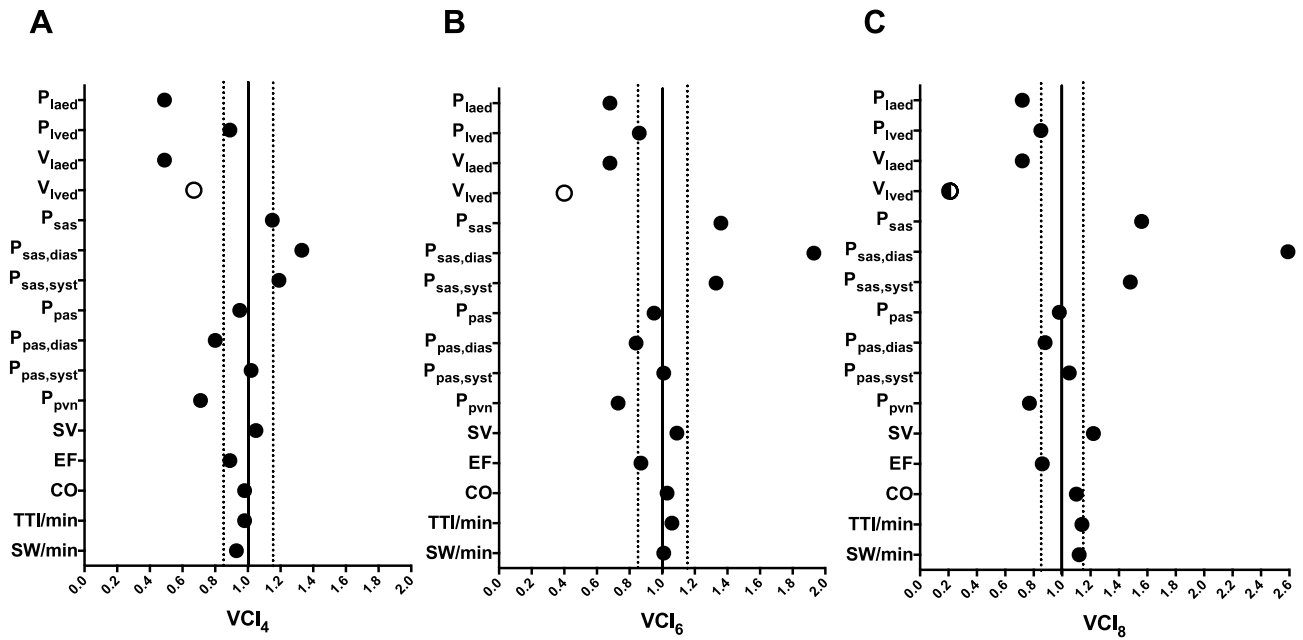

**Fig 3. Variation comparative index (VCI) graphs for computed cardiovascular variables.** (A) VCI$_4$, (B) VCI$_6$, (C) VCI$_8$. In each plot the solid line intercepts VCI axis at 1, while the two dashed lines intercept VCI axis at 0.85 and 1.15 (pre-defined cut-off between relevant, i.e. VCI ≥ 1.15 and ≤ 0.85, and not relevant variations, i.e. 1.15 > VCI > 0.85); with respect to the corresponding 1 MET simulation value, black dot indicates that the variable increases both in SHR and HHR exercise simulations, white dot indicates that the variable decreases both in SHR and HHR exercise simulations, black/white dot indicates that the variable undergoes an increase in SHR exercise simulation and a decrease in HHR exercise simulation.

doi:10.1371/journal.pone.0169967.g003

In this context, our computational work investigates how the cardiovascular response to exercise during AF is affected by different resting HR. The adopted mathematical model is validated in resting conditions [4] and represents an efficient tool to shed light on the relative changes in cardiovascular variables triggered by physical exertion in the setting of AF. The present work is computational, thus not to be considered definitive. Its goal is to characterize and understand some of the basic mechanisms underlying the hemodynamic changes, leaving aside the additional role of other factors, such as associated pathologies and/or drug effects. Given the few and conflicting data available, awaiting the actually missing systematic validation, the present exploratory model provides, in our opinion, significant hints into a relevant AF topic.

One of the pathophysiological features of AF is the loss of atrial systole ("atrial kick"), which constitutes the phase of active ventricular filling during the cardiac cycle. Consequently, a general reduction in ventricular filling is known to occur. However, the contribution of this cardiac phase strongly depends on HR: atrial systole becomes essential at faster HR, such as during physical exercise, since passive ventricular filling in early diastole is reduced due to a shortened diastolic time. In HHR simulations, characterized by faster HRs, the loss of atrial kick has a stronger influence on the heart and cardiovascular system, if compared to SHR simulations. In fact, V$_{lved}$ tends to remain constant in SHR simulations, while it drops in HHR simulations: a possible explanation is that while in SHR simulations the boosted venous return to the left heart during exercise counterbalances the impact of the loss of atrial systole, in HHR simulations this impact is magnified by the faster HR, preventing the venous return to compensate for the lack of atrial contraction and ultimately resulting in a decrease in V$_{lved}$.

The reduced V$_{lved}$ directly affects SV, which displays a greater increase in SHR simulations, with the computed VCI$_8$ reaching 1.22. This is reflected into a smaller increase in computed blood pressure values ($P_{sas}$, $P_{sas,syst}$, $P_{sas,dias}$) in HHR simulations, hinting that a permanent AF patient with a slower HR could have a more appropriate blood pressure response during exercise. Moreover, even if not reaching the formal limit of significance for VCI, a trend towards a less sustained increase in CO can be seen, with the computed $VCI_8$ for CO equaling 1.10. Fig 4A and 4B show mean CO values in SHR and HHR simulations and the correspondent VCIs.

Concomitantly, in HHR simulations the decreased $V_{lved}$ inversely affects left atrium hemodynamics, namely with an increase in $V_{laed}$ and $P_{laed}$. The upstream pulmonary circuit is influenced by left atrial overload: in fact, $P_{pvn}$, already starting from light effort, displays a greater increase in HHR simulations (Fig 4C and 4D), which is also reflected into a more sustained increase in $P_{pas,dias}$. Given the likelihood that an increase in pulmonary venous pressure leads to exertional dyspnea[32] and that a steeper rise in pulmonary venous pressure value has shown to relate to a reduced exercise tolerance, at least in selected populations[33], it is plausible that a permanent AF patient with a higher resting HR could suffer from dyspnea earlier during exercise than a patient with a lower one.

This interpretation of the data is further supported when comparing AF simulation results (Table 4) to the corresponding SR simulations (Table 2). In fact, as it arises comparing the two tables, AF SHR simulations values tend to be comparable to the corresponding SR simulations, differently from the HHR simulations. More in details, focusing on systolic systemic pressure ($P_{sas,syst}$), and pulmonary vein pressure ($P_{pvn}$), one can observe that the mean values at the same level of exercise are similar between SR and AF SHR simulations, while AF HHR results clearly deviate from these values. The combined comparison between SR and AF (SHR and HHR) simulations reveals, in fact, that, among the AF-induced features, higher HR mostly affects the AF hemodynamics under exertion. In fact, since SR and AF SHR simulations have the same mean HR and give similar results, loss of atrial kick and RR irregularity are not able themselves to promote significant variations. On the contrary, a consistent HR increase (as set in AF HHR) enhances AF-changes under exercise.

## Limitations

In addition to the aforementioned, the following limitations should be kept in mind. First, the present mathematical model does not predict the eventual impact that AF-associated pathologies (for example, hypertension) and rate control drugs (digoxin, beta blockers, non-dihydropyridine calcium channel blockers) could exert on the cardiovascular system. Second, the model does not account for the impact on peripheral resistances of the short-term regulation induced by the baroreceptor mechanisms. Moreover, as the passive minimum elastance is kept constant for all grades of exertion, the influence of altered lusitropy is not included. In addition, there is evidence that atrial compliance is reduced with AF[26, 27]. However, this aspect is not included in the present modelling since we checked that a proper decrease of the left and right atrial compliances (of the order of 10%) does not substantially alter the observed picture (output variations with respect to the reference results are well within 1%). Third, the coronary circle is not taken in account. Finally, the present study considered "relevant" variations, within the correspondent SHR and HHR simulations, with a VCI $\geq 1.15$ or $\leq 0.85$: concerning variables not reaching this limit, it cannot be totally excluded that their relative variation in response to a certain amount of exercise is not influenced by the resting HR, since the present modeling could not be able to detect relevant variations in those quantities.

## Conclusions

In conclusion, awaiting further clinical evidence, the present computational model suggests that a slower resting HR in permanent AF is associated with an improved hemodynamic response during exercise. In fact, pulmonary venous pressure undergoes a slighter increase and systemic blood pressure a more relevant increase upon exertion when the resting HR is slower. Present findings warrant clinical validation but suggest that, on a computational basis, targeting a slower resting HR might be preferable in permanent AF patients.

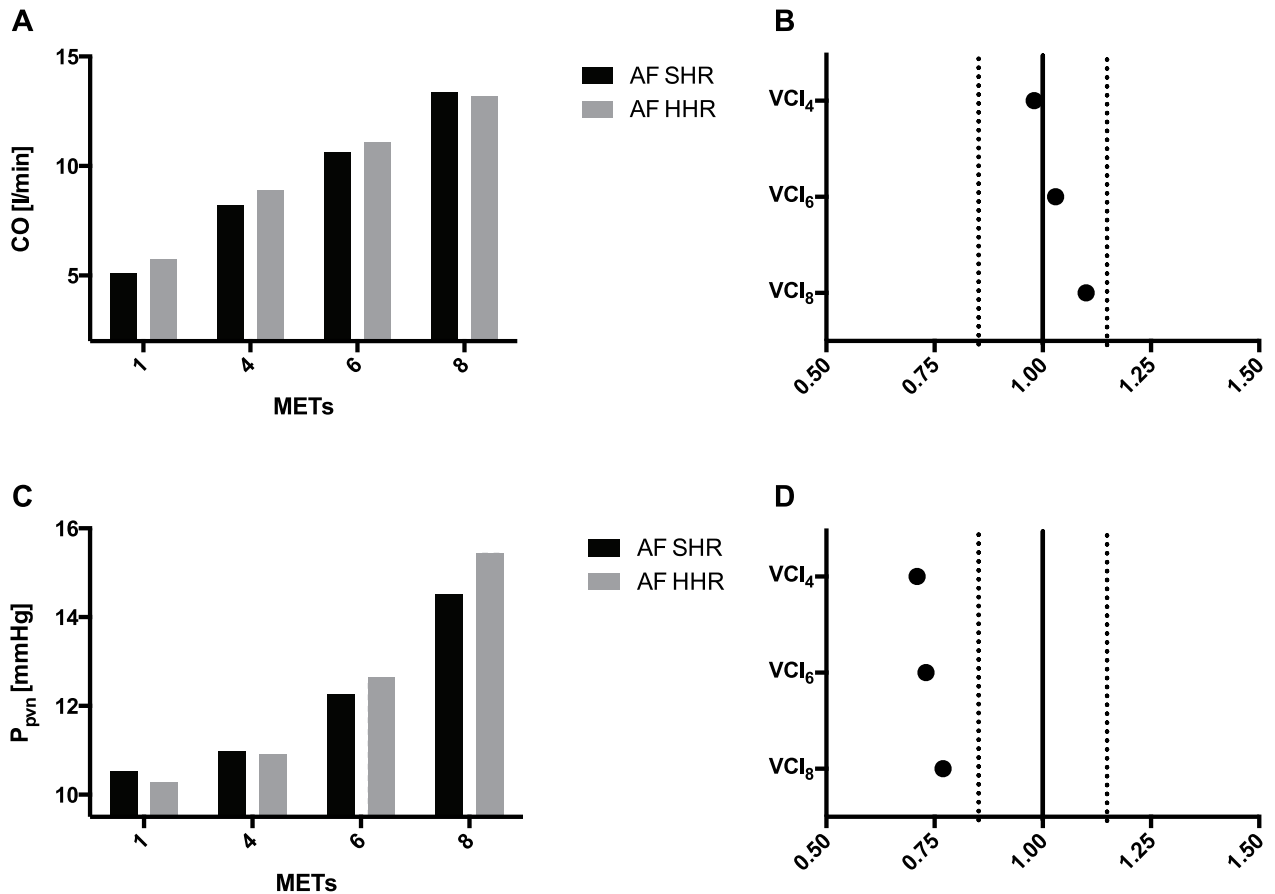

**Fig 4. Mean values and VCIs for cardiac output (CO) and pulmonary venous pressure (P_{pvn}).** (A) Mean CO as function of exercise intensity in SHR and HHR simulations. (B) Plot of the VCIs calculated for CO. (C) Mean P_{pvn} as function of exercise intensity in SRH and HHR simulations. (D) Plot of the VCIs calculated for P_{pvn}. In plot B and D, the solid line intercepts VCI axis at 1, while the two dashed lines intercept VCI axis at 0.85 and 1.15 (pre-defined cut-off between relevant, i.e. VCI $\geq$ 1.15 and $\leq$ 0.85, and not relevant variations, i.e. 1.15 > VCI > 0.85); with respect to 1 MET simulation value, black dot indicates that the variable increases both in SHR and HHR exercise simulations.

## Supporting Information

**S1 Table. Resting (1 MET) simulation parameters.** Lumped-model parameters at baseline. P: pressure [mmHg], V: volume [ml], Q: flow rate [ml/s], $\vartheta$: valve opening angle [rad]. E: elastance [mmHg/ml], C: compliance [ml/mmHg], R: resistance [mmHg s/ml], L: inductance [mmHg $s^2$/ml], CQ: flow coefficient [ml/(s mmHg$^{0.5}$)], K: valve coefficient. List of subscripts. la: left atrium, lv: left ventricle, ra: right atrium, rv: right ventricle, mi: mitral, ao: aortic, ti: tricuspid, po: pulmonary, un: unstressed, min: minimum, max: maximum, sas: systemic aortic sinus, sat: systemic artery, sar: systemic arterioles, scp: systemic capillary, svn: systemic vein, pas: pulmonary artery sinus, pat: pulmonary artery, par: pulmonary arterioles, pcp: pulmonary capillary, pvn: pulmonary vein, p: effect of pressure force, f: frictional action, b: velocity effect on the valve dynamics, due to blood motion, v: vortex effect on the valve dynamics. (DOCX)

**S2 Table. Sensitivity analysis of the AF simulations.** Cardiovascular parameters in Table 1 are varied, one at a time, by +5% and -5% for the two extreme conditions: 1 MET and 8 METs. Mean values of selected hemodynamic outputs ($P_{sas}$, $P_{pvn}$, $P_{pas,dias}$, SV) are reported in SHR. (DOCX)

## Author Contributions

**Conceptualization:** MA SS AS FG LR.

**Data curation:** SS LR.

**Formal analysis:** SS AS.

**Investigation:** MA SS AS FG LR.

**Methodology:** SS LR.

**Project administration:** MA LR.

**Resources:** SS LR.

**Software:** SS LR.

**Visualization:** SS AS.

**Writing – original draft:** MA SS AS FG LR.

**Writing – review & editing:** MA SS AS FG LR.